\documentstyle[pra,aps,twoside,multicol,rotate,epsf]{revtex}

\begin{document}
\draft

\title{Structure Determination of Disordered Metallic
Sub-Monolayers by Helium Scattering: A Theoretical and Experimental Study
}

\author{A.T. Yinnon,$^a$ D.A. Lidar (Hamburger),$^{a,b}$
R.B. Gerber,$^{a,c}$ P. Zeppenfeld,$^d$ M.A. Krzyzowski,$^d$ and
G. Comsa$^d$}
\address{
$^a$Department of Physical Chemistry and The Fritz Haber Center for Molecular
Dynamics, The Hebrew University of Jerusalem, Jerusalem 91904, Israel\\
$^b$Department of Chemistry, University of California - Berkeley, Berkeley, CA
94720, USA\\
$^c$Department of Chemistry, University of California - Irvine, Irvine, CA
92697, USA\\
$^d$Institut f\"{u}r Grenzfl\"{a}chenforschung und Vakuumphysik,
Forschungszentrum J\"{u}lich, Postfach 1913, D-52425 J\"{u}lich,
Germany
}

\maketitle

\begin{abstract}
An approach based on He scattering is used to develop an atomic-level
structural model for an epitaxially grown disordered sub-monolayer of
Ag on Pt(111) at 38K. Quantum scattering calculations are used
to fit structural models to the measured angular intensity distribution
of He atoms scattered from this system. The structure obtained corresponds
to narrowly size-dispersed compact clusters with modest translational
disorder, and not to fractals which might be expected due to the low
surface temperature. The clusters have up to two layers in height, the
lower one having few defects only. The relations between specific
features of the angular scattering distribution, and properties such
as the cluster sizes and shapes, the inter-cluster distance
distribution etc., are discussed. The results demonstrate the
usefulness of He scattering as a tool for unraveling new complex
surface phases.

\end{abstract}

\begin{multicols}{2}
\markboth{{Yinnon, Lidar, Gerber, Zeppenfeld, Krzyzowski and
Comsa}}{{\it Surf. Sci. Lett.} 1998}

\section{Introduction}

The physical and chemical properties of heteroepitaxial metallic
sub-monolayers depend on their structure, and the determination of the
latter is therefore of considerable technological and scientific
importance. Scanning tunneling microscopy (STM) and scattering
techniques have proved to be powerful methods for determining adlayer
structure
\cite{Comsa:book,Rieder:1,Ernst:2,Ibach,Kern:1,Lagally:STM,Comsa:STM1,Behm}.
The STM technique, for instance, has many advantages but is inherently
local and therefore the statistical characterization of disordered
surfaces requires averaging over many surface patches. In contrast, a
single atomic beam scattered from a surface can cover an entire
adlayer at once. Thermal He scattering enjoys several additional
advantages over other methods for identifying adlayer structure. (i)
The wavelength of a thermal He atom matches the size of metal surface
unit cells. (ii) Interference due to the quantum nature of the He
particles results in a high sensitivity to surface details. (iii) The
He atoms do not perturb the surface significantly. Finally, in
contrast to x-rays and neutrons which penetrate the bulk, He atoms
solely probe the topmost surface layer.\\
So far, little is known on the relation between different kinds of
adlayer disorder and the corresponding scattering patterns, which is
clearly necessary for the determination of structural properties from
experiment. Studies in this field include the case of defects at very
low concentration \cite{Lahee:Fr,Heller,Peppino,me:optical}, models of
translationally random small compact clusters
\cite{Jonsson2,me:Heptamers} and fractal surfaces
\cite{Pfeifer:fractals,me:fractals}. In this letter, we report a
detailed atomic-level structure determination based exclusively 
on He scattering. We studied He scattering data for an epitaxially
grown disordered Ag sub-monolayer on a Pt(111) surface at 38K. Using
quantum scattering calculations, various models of surface disorder
were compared to the experimental data. The sensitivity of the
scattering distributions to the structural features of the surface
allowed us to progressively narrow down the class of disorder, until a
satisfactory fit was obtained. A first step in this direction was
taken in Ref.\cite{me:Ag-systems}, where we showed that the broad
classes of (a) isolated, translationally random adatoms and (b)
fractal structures can be ruled out as plausible models of the surface
structure. On the other hand a qualitative agreement was found with
scattering intensities resulting from a model of translationally
random, narrow size-dispersed compact clusters. Here we report new
results revealing significant quantitative agreement between He
scattering simulations and the same experimental data, from which a
clear picture of the surface structure emerges. While the resulting
fit between scattering intensities is not unique, the procedure
introduced here for the analysis of He scattering data is at least
capable of ruling out several plausible models of surface
disorder. The structure identified in this procedure represents, in
our view, a new metastable phase in the sense that it is composed of
compact clusters with modest translational disorder, similar to that
seen, e.g., in detailed STM experiments, mainly for Pt/Pt(111)
\cite{Comsa:STM1}.

\section{Experimental Procedure}
The thermal He scattering experiments
were performed in an ultrahigh vacuum (UHV) system. The Pt(111) sample
was cleaned and characterized in situ and a sub-monolayer of Ag was
evaporated onto the 38K Pt(111) surface by means of a Knudsen
cell. The scattering apparatus was equipped with a time-of-flight
(TOF) spectrometer, allowing to separate the elastic from the
inelastic He scattering intensity. The scattered He atoms were
recorded as a function of the wavevector transfer $\Delta K$ parallel
to the surface. The outgoing He atoms were measured in the incidence
plane. Thus $\Delta K = k_i(\sin\theta_f - \sin\theta_i)$, where $k_i$
is the incident wavevector, and $\theta_i,\theta_f$ are the incident
and outgoing scattering angles, respectively. The total scattering
angle was held constant: $\theta_i + \theta_f = 90^{\circ}$.\\ {\it
Theoretical model:} Scattering calculations were performed using the
Sudden Approximation (SA) \cite{Benny:Sud1}, which has proven to be
very successful in describing scattering from both ordered
\cite{Benny:4} and disordered surfaces \cite{Benny:XeKr}, where the
very large grids required prevent the use of numerically exact
techniques. The SA has been reviewed extensively \cite{Benny:review2}:
it assumes that the momentum transfer parallel to the surface is small
compared with that normal to the surface, i.e., $|\Delta K| \ll 2k_z$,
where $k_z$ is the wavevector component perpendicular to the surface.
On the basis of the experience gained with the SA, including tests
against numerically exact calculations
\cite{Benny:Sud1,Benny:4}, we estimate that the main predictions of
the SA should be reliable for the systems considered here. The He-Pt
interaction was modeled using the laterally averaged potential of
Ref.\cite{Zeppenfeld:2} which was extracted from experimentally
determined surface resonances on Pt(110). We assume that the
He/Pt(111) potential is smooth along the surface. Other experimental
and theoretical studies of the He/Pt(111) system support this to be an
excellent approximation. A further assumption in this work is that the
influence of surface vibrations on the elastic angular intensity
distribution is small, so they can be neglected.\\
The He/Ag adlayer interactions were represented by a sum of pairwise
potentials between He and isolated, adsorbed Ag atoms. Clearly, on
clustering of Ag adatoms, some rearrangement of the local electron
density is to be expected. Still, we believe a pairwise potential to
be a reasonable approximation, especially for small adatom
islands. These seem to be present here based also on STM data
\cite{Roder}. The He/Ag adatom interaction was determined by fitting
calculated SA single-adatom cross sections to experimental data. Our
fitted potential accurately reproduces the experimental cross sections
over a large range of incidence energies
\cite{me:Ag-systems}.

\section{Results}
\begin{figure}
\hspace{-3em}
\vspace{-1.0em}
\epsfysize=7cm
\epsfxsize=10cm 
\epsffile{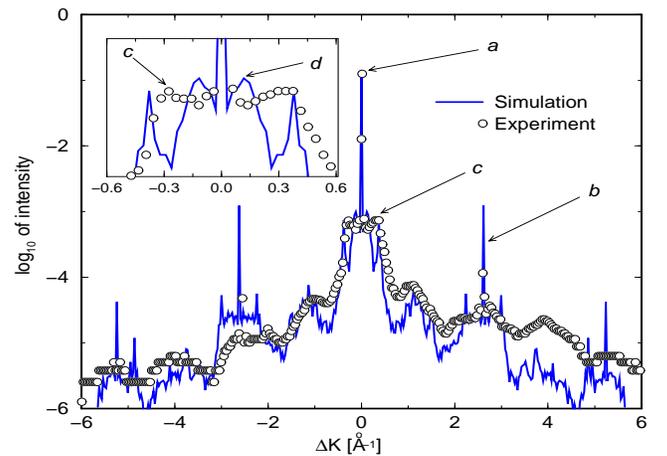}
\caption{Normalized, purely elastic scattering along the [11\={2}]
\protect\newline azimuth  recorded from the Ag/Pt(111) system at 50\%
coverage \protect\newline and a surface temperature of 38K. Intensity
normalized to that \protect\newline of incoming beam. \hfill \hspace{6.0cm}}
\label{fig:TvsE}
\end{figure}

\begin{figure}
\hspace{-6em}
\vspace{-2.5em}
\epsfysize=6cm
\epsfxsize=12cm
\epsffile{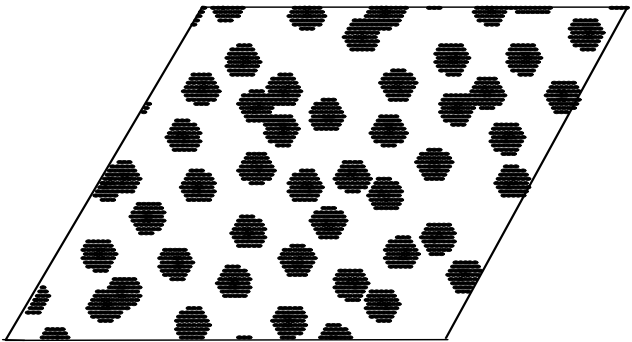}
\caption{Typical configuration of Ag islands on Pt(111). \hfill
\hspace{10.0cm}}
\label{fig:islands}
\end{figure}

Our main experimental result is presented in
Fig. \ref{fig:TvsE}: it shows the purely elastic scattering
intensities for a He beam with $k_i$=6.43${\rm \AA}^{-1}$=
21.6meV impinging 
upon a Pt(111) surface covered with 0.5 monolayers of Ag. The
orientation of the scattering plane is along the
[11\={2}]-direction. The main theoretical result is represented by the
solid line: the calculated scattering intensities from the structure
which we found to produce the best fit with the experimental
data. Before going into details we present the central features of
this structure:\\
Nearly hexagonal islands with few atomic edge defects, of side-lengths
3-5 unit cells along the [11\={2}]-direction, and a height of up to
two layers. Such hexagons can accommodate 50-60 atoms and we find the
average number of atomic defects per island to be around 10. At a
coverage of 0.5 monolayer about 65 islands are formed per 100$^2$ unit
cells. Whereas our analysis can rule out a third layer, it is not
sensitive to the detailed structure of the second layer. The existence
of a second layer is corroborated by the dependence of in- and
out-of-phase scattering on the incidence energy
\cite{Krzyzowski:thesis}. The average number of second layer atoms per
island can be inferred to be about 30 from the numbers above. The
positional distribution of the island centers deviates by no more than
three Pt(111) lattice constants per island from a hexagonal
superlattice with a lattice constant of 12-13 unit cells. A typical
configuration of the first layer is shown in
Fig. \ref{fig:islands}. Our analysis cannot exclude island-shapes
differing slightly from hexagonal. However, the existence of compact
clusters is supported by STM experiments at low temperatures
\cite{Roder}. The hexagonal shape is the energetically preferred one
in such a system, although it must be kept in mind that equilibrium
considerations need not necessarily prevail here. The main conclusion of our
study is that the Ag layer produced in the experiment has the
structure described above. It is remarkable that a {\em fractal} phase
can be ruled out, as one may have expected fractals due to the low adatom
mobility at our low surface temperature. A fractal surface would
result in a {\em smooth} decay of the off-specular intensity
\cite{me:Ag-systems}, in complete discrepancy with the observed
intensity pattern (Fig. \ref{fig:TvsE}). Having stated the structural
characteristics of the surface, we turn to a presentation of the
analysis which led us to our conclusions, and to a discussion of the
sensitivity and uniqueness of our fit.

\noindent Features in the experimental
angular intensity distribution due to {\em coherent} scattering can be
assigned directly, whereas those due to various incoherent scattering
mechanisms require comparison to simulation results of scattering from
different types of disorder. We consider the following features:\\
{\bf Specular Peak:} This feature contains information about island
{\em corrugation} and {\em size}. The relatively large specular {\em
intensity} ($\sim\! 0.1$, Fig. \ref{fig:TvsE}, arrow {\it a}) is
typical of He scattering from a surface with large flat patches
\cite{Comsa:layers}. This serves
as an indication of the smoothness of the Ag adlayer. Furthermore,
off-specular scattering from metallic surfaces results mostly from
collisions of He with the island-edge region \cite{me:fractals}. Thus
the large specular peak provides evidence for {\em compact} Ag
islands, several atoms in diameter. In contrast, the {\em width} of
the specular peak is determined by the average diameter $d$ of the
islands \cite{me:Ag-systems}. The observed width of
0.6${\rm \AA}^{-1}$ results in 
$d\approx$21${\rm \AA}$, or 8-9 Pt unit cells along the
[11\={2}]-direction [the Pt(111) lattice constant being $a$=2.77${\rm
\AA}$], consistent with the above requirement of large 
islands. The simulation reproduces both the intensity and the width of
the specular peak with remarkable accuracy.

\vspace{-1.5em}

\begin{figure}
\hspace{-3em}
\vspace{-1.0em}
\epsfysize=7cm
\epsfxsize=10cm
\epsffile{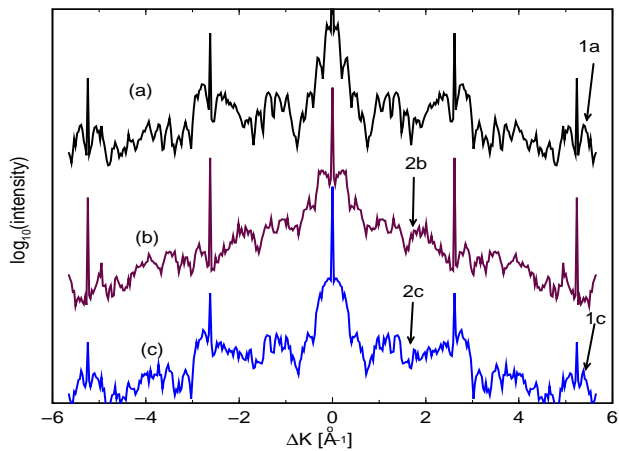}
\caption{Calculated He scattering intensities for various test
\protect\newline systems of Ag hexagons on Pt(111). (a) 91 Ag atoms
per \protect\newline hexagon; (b) 19 Ag atoms; (c) 19 or 91
atoms. \hfill \hspace{6.0cm}}
\label{fig:checks}
\end{figure}

\noindent
{\bf Bragg Peaks:} Strong Bragg interference peaks can be
observed at $\Delta K\!=\!-2.54{\rm \AA}^{-1},2.60{\rm \AA}^{-1}$
(Fig. \ref{fig:TvsE}, arrow {\it b}). Although growth of the first Ag
layer on Pt(111) is known to be pseudomorphic in the entire
temperature range \cite{Krzyzowski:thesis}, another interpretation
consistent with the data is that the average $|\Delta
K|\!=\!2.57{\rm \AA}^{-1}$ is the result of an {\em effective unit
cell length} of $4\pi/(2.57\sqrt{3})=2.82{\rm \AA}$, in between the
Pt(111) and Ag(111) values of 2.77${\rm \AA}$ and
2.89${\rm \AA}$ respectively. Indeed, strain does not permit the Ag
islands to relax completely to their natural unit cell size. The
absence of Pt(111) Bragg peaks indicates that the atomic corrugation
of the island surface is much larger than that of the underlying
Pt(111) layer. In our simulations only the island Bragg peaks are
reproduced, as the Pt(111) surface is treated as completely flat. The
discrepancy in intensities probably indicates that the pairwise
additive potential is too repulsive in this case. Assignment of
further features requires comparison with simulation.\\
{\bf Interference Maxima at $\Delta K\!=\!\pm 0.3{\rm \AA}^{-1}$}:
Indicated by arrow {\it c} in Fig. \ref{fig:TvsE} and shown enlarged
in the insert, this feature, unlike other interference peaks, is
nearly independent of azimuthal direction, but does depend on surface
temperature and the adsorption rate. It results from an interference
between adjacent island {\em edges} and can thus be traced to a
corresponding average {\em inter-island center-to-center distance} of
$l \sim 30{\rm \AA}$. The scattering calculations are extremely
sensitive to this distance, along with the average island diameter
$d$. As can be seen in Fig. \ref{fig:TvsE}, after extensive
fine-tuning of the parameters $l$ and $d$, the simulation results are
compatible with this feature. This fit is one of the main sources of
our confidence in the present structural determination, as the
combination of $l$ and $d$ poses severe constraints on the allowed
geometries at such high coverage. These constraints indicate almost directly an
important structural aspect: The {\em modest} amount of translational
disorder.\\
{\bf Other Off-Specular Peaks}: A detailed theoretical analysis of the
off-specular structure was given in Ref.\cite{me:Ag-systems}. Some of
the peaks can be assigned to either Fraunhofer \cite{Lahee:Fr} or
rainbow \cite{me:Heptamers} scattering. These, in turn, can be traced
back to the ``form factor'' due to scattering from an {\em isolated}
island. An example is the peak indicated by arrow {\it d} in the
insert of Fig. \ref{fig:TvsE}, which is probably a rainbow effect due
to a single island. If in reality the electronic density shape depends
on cluster size this is expected to be ``washed out'' in the
experiment, as observed here. A great deal
can be learned by a trial and error process
attempting to assign peaks to islands of different sizes. One
parameter which can be determined in this way is the island
concentration. For a system consisting of isolated and randomly
positioned {\em identical} islands the intensity of its characteristic
off-specular peaks is proportional to the concentration
\cite{me:Ag-systems}. However, when there is a distribution of island
shapes and sizes, as in Fig. \ref{fig:islands}, each island
contributes its own peaks and the simple proportionality is
lost. Using the trial and error approach, we arrived at a
concentration of 65 islands per 100$^2$ unit cells. As can be seen in
Fig. \ref{fig:TvsE}, this yields a quantitative agreement for $\Delta
K < 2{\rm \AA}^{-1}$ between experiment and simulation, in peak
positions, widths and intensities. The agreement is less quantitative
for the intensities at $\Delta K > 2{\rm \AA}^{-1}$, but a
qualitative agreement remains, in that the positions of all minima and
maxima are correctly reproduced. It should be mentioned that a basic
flaw of the SA, in disagreement with experiment, is that it is
symmetric in $\pm \Delta K$, regardless of the incidence angle. The SA
inherently assumes small parallel momentum transfer \cite{Benny:Sud1},
to which the discrepancy for high $\Delta K$ can be
attributed. Nevertheless, the agreement is noteworthy for all $\Delta
K$, and highly sensitive to the parameters of the surface structure.\\
{\bf Uniqueness of the Structural Model}: The scope of this letter
will not permit us to convey in detail the sensitivity of the fit
shown in Fig. \ref{fig:TvsE}. Essentially the only parameters of the
adlayer that do {\em not} have a substantial effect on the calculated
scattering intensities are the structure of the second layer and the
detailed internal structure of the hexagons. However, all other
parameters, and in particular the average island diameter and
polydispersivity in sizes, average distance and positional
distribution, concentration and number of defects, strongly affect the
quality of the fit and can be determined with a high degree of
confidence by the present trial and error approach. Of course, our
search was not entirely random but guided by insight as to which
geometric structures can produce a certain feature. To illustrate this
and the sensitivity of the intensity distribution to system
parameters, consider some features of Fig. \ref{fig:checks}. Shown there are
the calculated intensities for random distributions of non-overlapping
perfect hexagons of (a) 91 Ag atoms, (b) 19 Ag atoms, (c) 19 or 91 Ag
atoms (equal number of islands), all at 15\% coverage. Arrow 1a
indicates a peak which is characteristic of the 91-atom hexagons (a),
and is absent in the intensity spectrum of the 19-atom hexagons
(b). However, it is clearly present in the 
spectrum of the combined system, as indicated by arrow 1c. Similarly,
arrow 2b points at a feature which is present for 19-atom hexagons but
not for 91-atom hexagons, yet is present in the spectrum of the
combined system, as shown by arrow 2c. The reader will easily
recognize additional features in Fig. \ref{fig:checks}(c) which can be
attributed uniquely to only one of the systems. In a manner similar to
this we have been able to {\em resolve} the experimental intensity
distribution of Fig. \ref{fig:TvsE} and conclude that its features are
due to the structure shown in Fig. \ref{fig:islands}.

\section{Conclusions and Outlook}
In conclusion, this study 
demonstrated that He scattering is
capable of performing a ``crystallography'' of disordered surfaces. We
reported one of the first 
theoretical-experimental detailed atomic-level structure determinations
of a disordered surface layer by He scattering: A well-defined geometry
of narrowly size-dispersed, compact hexagonal
clusters, with modest translational disorder, is formed by Ag
deposited on Pt(111) at 38K. This geometry comes as a surprise, since
it has been 
shown using variable-temperature STM that at 35K individual Ag adatoms
do not diffuse on the Pt(111) 
surface on a time scale of at least two hours \cite{Brune:95}. Our
results clearly necessitate some mechanism for a significant
lowering of the diffusion barrier. How this comes about must be
contained in electronic structure 
considerations, which we do not know at present. Why the high degree
of translational order? We
speculate that there are long-range forces between large
clusters, and these tend to order the clusters. The possible origins
of such forces are electrostatic 
(polarization) interactions, or long-range elastic interactions
\cite{Zeppenfeld:95a}. A detailed study of these issues will be the
subject of future study.

\section*{Acknowledgements}
This work was supported by Grant
No. I-215-006.5/91 from the German-Israel 
Foundation for Scientific Research (G.I.F.) to R.B.G. and G.C. The research
was supported in part by the Institute of Surface and Interface Science at
U.C. Irvine. We would like to thank Dr. I. Farbman and Prof. O. Biham for
helpful discussions.

\end{multicols}

\end{document}